\documentstyle{article}

\newcommand{\beq}{\begin{equation}}
\newcommand{\eeq}{\end{equation}}
\newcommand{\beqa}{\begin{eqnarray}}
\newcommand{\eeqa}{\end{eqnarray}}

\def\ra{\rightarrow}
\def\x{\times}
\def\etal{\it etal,}

\newcommand{\NC}[3]{{\elevenit Nuo. Cim.} {\elevenbf #1}, #2 (19#3)}
\newcommand{\RMP}[3]{{\elevenit Rev. Mod. Phys.} {\elevenbf #1}, #2 (19#3)}
\newcommand{\PR}[3]{{\elevenit Phys. Rev.} {\elevenbf #1}, #2 (19#3)}
\newcommand{\PL}[3]{{\elevenit Phys. Lett.} {\elevenbf #1}, #2 (19#3)}
\newcommand{\Rep}[3]{{\elevenit Phys. Rep.} {\elevenbf #1}, #2 (19#3)}

\newcommand{\PRL}[3]{{\elevenit Phys. Rev. Lett.} {\elevenbf #1}, #2 (19#3)}
\newcommand{\NP}[3]{{\elevenit Nucl. Phys.} {\elevenbf #1}, #2 (19#3)}
\newcommand{\con}[3]{{\elevenbf #1}, #2 (19#3)}

\def\Ev{\; {\rm eV} }
\def\ev{\; {\rm eV} }
\def\eV{\; {\rm eV} }
\def\gev{\; {\rm GeV} }
\def\etal{ {\it et al}.}
\font\tenbf=cmbx10
\font\tenrm=cmr10
\font\tenit=cmti10
\font\elevenbf=cmbx10 scaled\magstep 1
\font\elevenrm=cmr10 scaled\magstep 1
\font\elevenit=cmti10 scaled\magstep 1

\renewenvironment{thebibliography}[1]
 { \elevenrm
   \begin{list}{\arabic{enumi}.}
    {\usecounter{enumi} \setlength{\parsep}{0pt}
     \setlength{\itemsep}{3pt} \settowidth{\labelwidth}{#1.}
     \sloppy
    }}{\end{list}}

\begin{document}
\vglue 1.0cm
\begin{flushright}
\today \\
UPR-0552T\\
\end{flushright}
\vglue 2.0cm
\begin{center}{{\tenbf SOLAR NEUTRINOS}\footnote{Invited talk
presented at {\it Unified Symmetry in the Small and in the Large},
Coral Gables, Florida, January, 1993.}\\
\vglue 1.5cm  
{\tenrm PAUL LANGACKER\\}
\vglue 0.6cm
\baselineskip=13pt
{\tenit University of Pennsylvania \\ Department of Physics
\\ Philadelphia, Pennsylvania, USA 19104-6396\\}
\vglue 0.8cm
{\tenrm ABSTRACT}}
\end{center}
\vglue 0.3cm
{\rightskip=3pc
 \leftskip=3pc
 \tenrm\baselineskip=12pt
 \noindent
The status of solar neutrino experiments and their implications
for both nonstandard astrophysics ({\it e.g.,} cool sun models)
and nonstandard neutrino properties ({\it e.g.,} MSW conversions) are
discussed. Assuming that all of the experiments are correct, the
relative rates observed by Kamiokande and Homestake
are hard to account for by a purely astrophysical solution, while
MSW conversions can describe all of the data.
Assuming the standard solar model, there are two allowed regions for
MSW conversions into $\nu_\mu$ or $\nu_\tau$, with the non-adiabatic
solution giving a better fit than the large angle. For conversions
into sterile neutrinos there is only a nonadiabatic solution.
Allowing both MSW conversions and nonstandard astrophysics,
the data simultaneously
determine the temperature of the core of the sun to within five
percent, consistent with the standard solar model prediction.
The implications of the atmospheric $\nu_\mu/\nu_e$ ratio
and of a hot component of the dark matter are
briefly discussed, and the expectations of theoretical models motivated
by grand unification are summarized.
}
\newpage   
{\elevenbf\noindent 1. Introduction}
\vglue 0.2cm
\baselineskip=14pt
\elevenrm
We now have a good idea of the number of neutrinos.  From the $Z$
width $\Gamma_Z$ measured at LEP~\cite{lep} one has $N_\nu = 3.04
\pm 0.04$, where $N_\nu$ is the number of active (normal)
neutrinos of mass less than $M_Z/2$.  There is a complementary
limit $N_\nu' < 3.3$ from nucleosynthesis.  This applies only to
neutrinos of masses less than about 20~MeV \cite{nuc}, but
includes sterile neutrinos (neutrinos that do not have
interactions except for mixing) for a wide range of masses and
mixings~\cite{sterile}.

We still do not know whether any of the neutrinos have mass.
There is no compelling evidence for nonzero masses from
laboratory experiments. However, most new physics beyond the
standard model predicts $m_\nu \neq 0$ at some level.  Usually,
$m_\nu \sim v^2/M$ where $v$ is the weak interaction scale and
$M$ is the scale of new physics~\cite{seesaw}.  Therefore, small
neutrino masses are a probe of physics at very high scales.

On the other hand, there are indications of possible neutrino
masses from non-laboratory experiments.  The deficit of solar
neutrinos as observed by Homestake~\cite{homestake},
Kamiokande~\cite{kam}, GALLEX~\cite{gallex}, and SAGE~\cite{sage}
suggests that either there is nonstandard astrophysics or that
there are new properties of the neutrinos, such as nonzero
masses.  An overall deficit could be accounted for by a
lower temperature of the core of sun than is usually predicted.
However, the relative rates of the Homestake and Kamiokande
experiments strongly disfavor an astrophysical solution,
suggesting that there are new properties of the neutrinos
(assuming that the experiments are correct
within the stated uncertainties).

Similarly,
the deficit in the ratio of atmospheric
$\nu_\mu/\nu_e$ as
observed by the Kamiokande \cite{8a} and IMB \cite{imb}
collaborations in their contained events
suggests the possibility of neutrino oscillations~\cite{penn}.
No anomaly is
observed in the stopping or through-going upward muons.  However,
these are more sensitive to absolute theoretical calculations of
the fluxes than is the $\nu_\mu/\nu_e$ ratio.

Finally, there is still the possibility of a hot component
of the dark matter of the universe.  For $m_{\nu_\tau} \sim (1-28)
\Ev$ the tau neutrino would be important.  Smaller masses would be
cosmologically irrelevant, while larger ones would over-close the
universe.  Hot dark matter has long been out of favor as the only
component of dark matter because it is difficult to explain how small
structures such as galaxies could have formed in the time available.
One would therefore need to combine the hot dark matter with other
ingredients such as cosmic strings, cold dark matter, or decaying
neutrinos to seed galaxies.  Such a cocktail has been suggested by a
number of authors as one explanation of the COBE and galaxy distribution
data~\cite{cobe}.

\begin{figure}
\vspace{5cm}
\caption{Neutrino oscillation parameters excluded or suggested by
various observations.  The combined MSW solar neutrino solution and
the region suggested by the atmospheric neutrinos are indicated.  The
predictions of various theoretical models are also shown.}
\label{fig1}
\end{figure}

\vglue 0.6cm
{\elevenbf\noindent 2. Solar Neutrinos}
\vglue 0.2cm
The predictions for the fluxes in the Homestake, Kamiokande, and
gallium experiments from two recent theoretical studies are shown in
Table~\ref{tab1}.  There is reasonable agreement between them,
especially for the gallium experiments.  However, the
Bahcall-Pinsonneault (PB) \cite{bp} calculation predicts a somewhat
higher $\,^8B$ flux than that of Turck-Chi\`{e}ze (TC) \cite{tc}.

\begin{table} \centering
\begin{tabular}{|c|c|c|} \hline
Theory & SSM (BP) & SSM (TC) \\ \hline
Homestake (Cl) & $8 \pm 1 $ SNU & $6.4 \pm 1.3$ SNU \\
Kamiokande & $1 \pm 0.14$ (arb units) & $0.77 \pm 0.20$ \\
gallium & $132 \pm 7 $ SNU & $125 \pm 5 $ SNU \\ \hline
\end{tabular}
\caption{Predictions of Bahcall-Pinsonneault~(BP)~\protect\cite{bp}
and Turck-Chi\`{e}ze~(TC)~\protect\cite{tc} for the solar neutrino
fluxes.  All uncertainties are at one standard deviation, which for BP
is defined simply as (total range)/3.}
\label{tab1}
\end{table}

The current experimental results are compared with the theoretical
predictions in Table~\ref{tab2}.  In each case there is a significant
reduction from the predictions of both standard solar
models.\footnote{ The Homestake experiment has also seen
evidence of a possible time dependence in their signal.  In the following this
will be ignored, because it has not been observed by the other
experiments and is not statistically compelling.  One might, however,
wonder whether this apparent time dependence affects the long term
average.  In fact, the periods 1970 --- 1977 and 1978 --- 1990 yield
almost identical averages: 2.1 and 2.2 SNU, respectively.}

\begin{table} \centering
\begin{small}
\begin{tabular}{|c|c|c|c|} \hline
 & Rate & Rate/SSM (BP) & Rate/SSM (TC) \\ \hline
Homestake & $2.1 \pm 0.3$ SNU & $0.26 \pm 0.04$ & $0.33 \pm
0.05$\\
Kam-II (1040 days) &  & $0.47 \pm 0.05 \pm 0.06 $   & \\
Kam-III (395 days) &  & $0.56 \pm 0.07 \pm 0.06 $  & \\
Kam-II + III      &  & $0.50 \pm 0.07 $ & $0.65 \pm 0.09$ \\
\raisebox{.9ex}{(prelim syst.)}&  & & \\
GALLEX & $83 \pm 19 \pm 8 $ SNU & $0.63 \pm 0.14$ & $0.67 \pm
0.15 $ \\
SAGE (90 + 91) & $58^{+17}_{-24} \pm 14$ SNU  & $0.44 \pm 0.19$ &
$0.47 \pm 0.20$ \\
GALLEX + SAGE & $71 \pm 15$ SNU & $0.54 \pm 0.11$ & $0.57 \pm
0.12$ \\
\hline
\end{tabular}
\end{small}
\caption{The observed rates, and the rates relative to the standard
solar model (SSM) calculations of BP and TC.}
\label{tab2}
\end{table}

There are several resolutions of the discrepancy.  The first is that
something is wrong with the astrophysics.  Here one must distinguish
between the standard solar models, which are in excellent agreement
with all other data but which are far from the observed neutrino
fluxes, and nonstandard solar models which involve totally new
ingredients \cite{BB,BHKL,SS}.  Another possibility is that there is
something wrong with our understanding of the neutrinos.  To my mind
the simplest and most likely possibility is that of MSW
conversion~\cite{MSW},
which can be considered both within the SSM and within
nonstandard solar models \cite{BHKL,SS,SSB,GR,18a}.  One can also
consider MSW oscillations with three flavors~\cite{HKP}, vacuum
oscillations~\cite{APP,BPW,18a}, or more exotic possibilities, such as
the interesting suggestion of gravitational effects associated with a
violation of the principle of equivalence~\cite{gasp,HLP}, neutrino
decay~\cite{AJPP}, or large magnetic moments~\cite{25a}.  The latter
would probably be needed if there is indeed a time dependence, but
sufficiently large moments are apparently in conflict with limits from
red giants~\cite{mag}.  A final possibility is that one or more of the
experiments is incorrect or has significantly underestimated its
uncertainties.  For example, if one discounted the Homestake result
astrophysical solutions or larger ranges for $\Delta m^2$ would be
allowed.  However, I have no reason to doubt any of the experiments,
so I will restrict my comments to astrophysical solutions and to the
simple MSW solution.

\vglue 0.6cm
{\elevenbf\noindent 3. Astrophysical Solution}
\vglue 0.2cm
The standard model of Bahcall and Pinnsoneault~\cite{bp} incorporates
helium diffusion, new $S_{17}$ values, is in good agreement with other
calculations when the same parameters are used, and it is in
reasonable agreement with helioseismology and other observations.  The
uncertainties in the SSM have been estimated using Monte Carlo methods
by Bahcall and Ulrich~\cite{bu}, who considered uncertainties in the
parameters relevant to the opacities, nuclear cross-sections, {\it
etc.} As can be seen in Table~\ref{tab1}, the standard solar model is
not in agreement with the data for any reasonable range of the
uncertainties, and is therefore excluded.

Still possible, however, is some type of
nonstandard solar model (NSSM), which
may differ from the SSM by entirely new physics inputs such as weakly
interacting massive particles (WIMPs), a large core magnetic field,
core rotation, {\it etc.} \ Most of these models affect the solar
neutrinos by leading to a lower temperature of the core of the sun,
{\it i.e.}, $T_c<1$, where $T_c = 1$
corresponds to the standard solar model.  Many of these models are
rather {\it ad hoc}.  More important, all reasonable models lead to a
larger suppression of the Kamiokande counting rate, which is
essentially all $\,^8B$, than that of Homestake, which has
in addition a nontrivial component of $\,^7Be$ neutrinos.  This is in
contrast to the experimental data, and therefore almost all
nonstandard solar models are excluded unless one or more of the
experiments is incorrect.

This can be argued quantitatively,
following the treatment
of Bludman, Hata, Kennedy and myself
\cite{BHKL}.  For this, and also for the following discussion of
the astrophysical uncertainties within the MSW solution, I will use a
simplified treatment of the uncertainties.
The many astrophysical effects are parameterized
by an arbitrary core temperature $T_c$, with the standard solar model
corresponding to $T_c = 1 \pm 0.006$.  Nuclear physics uncertainties in
the production and detection cross-sections are also included.  This
simplified error treatment reproduces quite well the uncertainties in
the SSM and generalizes to the NSSM.

\vglue 0.2cm
{\elevenit\noindent 3.1. Cool Sun Models}
\vglue 0.1cm
In cool sun models we assume that because
of some new physics input the core temperature can be considerably
below the range $T_c = 1 \pm 0.006$ of the standard solar model.
Following Bahcall and Ulrich~\cite{bu} the temperature dependence of
the dominant flux components is $\varphi (\,^8B) \sim
T^{18}_c$, $\varphi(\,^7Be) \sim T_c^8$, and $\varphi (pp) \sim
T_c^{-1.2}$.  Actually, these are derived assuming small variations
around the standard solar model.  If one allows for much larger
deviations, as will be needed to account for the observed deficits,
these exponents can only be regarded as qualitative.  The most
questionable is the $pp$ rate.  The exponent of $-1.2$ is not a good
approximation when one gets very far away from the SSM.  It is more
realistic to assume that the $pp$ flux is reduced by a factor $f(pp)$
which is chosen so that the total solar luminosity remains constant.

The expected counting rates $R$ for
each experiment relative to the expectations of the standard solar
model are then
\beqa R_{c\ell} = &0.26 \pm 0.04 = & ( 1 \pm 0.033)  [ 0.775 (1
\pm 0.10) T_c^{18}
 + 0.150 ( 1 \pm 0.036) T_c^8 \nonumber \\ &&
 + \;{\rm small}\;] \nonumber \\
R_{\rm Kam} = & 0.50 \pm 0.07 = & (1 \pm 0.10) T_c^{18} \nonumber \\
R_{\rm Ga} = & 0.54 \pm 0.11 = & ( 1 \pm 0.04) [ 0.538 ( 1 \pm
0.0022)
f(pp) \nonumber \\ && + 0.271 (1 \pm 0.036) T_c^8
 + 0.105 (1 \pm 0.10) T_c^{18} + \; {\rm small}\, ]. \label{eqa} \eeqa
In (\ref{eqa}) the temperature dependence of the individual flux
components is displayed.  The uncertainties which multiply the
overall rates are from the nuclear detection
cross-sections, and those which multiply the individual
flux components are cross-section uncertainties for the relevant
reactions in the sun.  The latter must, of course, be correlated from
experiment to experiment.

The best fit is for $T_c = 0.92 \pm 0.01$.  This requires
an enormous deviation from the SSM.  Even more disturbing is that it
is a terrible fit: $\chi^2 = 20.6$ for $2\; d.f.$, which is
statistically excluded at the $99.9\% \,cl$.  If we accept the
experimental values, a cool sun model simply cannot account for the
data, because one expects $R_{\rm Kam} < R_{\rm cl}$, contrary to what
is seen \cite{BHKL}.

This conclusion is much more general than the specific exponents
assumed above.  For example, if one simply assumes that $\varphi
(\,^8B) \sim T_c^{n_B}$ and $\varphi (\,^7 (Be) \sim T_c^{n_{Be}}$
then for any reasonable values of the exponents ({\it e.g.,}
$n_B = 27$, $n_{Be} = 15$, or
even arbitrary exponents satisfying $n_{Be}
\leq n_B$)
one again finds that
the data cannot be fit.
We have also found
\cite{BHKL} that even if one doubles all of the uncertainties in the
nuclear cross-sections one is not able to describe the data, because
the cross-section uncertainties are strongly correlated from
experiment to experiment.  Other NSSMs, such as models with an ad hoc
reduction of the $\,^8B$ flux \cite{BB} or explicit WIMP and
low $Z$ models, also fail for similar reasons.  The
conclusion is that as long as the stated experimental uncertainties
are taken seriously the nonstandard solar models are not a likely
explanation of the observations.  They would, however, be possible if the true
flux that should be observed in the Homestake experiment were $\geq 3
$ SNU (as opposed to $2.1 \pm 0.3$), but only if one  significantly
stretches the astrophysical uncertainties \cite{SS}.

\vglue 0.6cm
{\elevenbf\noindent 4. MSW Conversions}
\vglue 0.2cm
{\elevenit\noindent 4.1. Standard Solar Model}
\vglue 0.1cm
There have been a number of recent studies of the MSW \cite{MSW}
solution \cite{BHKL,SS,SSB,GR,18a}.  Here I will follow the work in
\cite{BHKL}, which uses the Bahcall-Pisonneault \cite{bp} production
distribution $\varphi_i (r)$ (for the $\rm i^{\rm th}$ flux component
as a function of the distance from the center), electron density
$n_e(r)$ (needed to calculate the MSW conversion), and neutron
density $n_n(r)$ (which enters for oscillations
$\nu_e \ra \nu_s$ into sterile neutrinos).  We have carefully
considered theoretical uncertainties; in particular, we have
parameterized the SSM uncertainties using the core temperature $T_c =
1 \pm 0.006$ relative to the SSM, which is due namely to uncertainties
in opacities, plus the production and detection cross-section
uncertainties as described earlier.  These agree well with the stated
uncertainties calculated by Bahcall and Pisonneault for the standard
solar model.  There are additional theoretical uncertainties involved
in the MSW conversion itself, including the uncertainties in $\varphi_i
(r)$ and $n_e (r)$.  We have made various estimates of these, which
turn out to be small.

For Kamiokande the detector resolution, threshold corrections, and
neutral current scattering (for the case of conversion into $\nu_\mu$
or $\nu_\tau$) have been included.  The effect of reconversion in the
earth has not yet been incorporated.  This will slightly affect the shape
of the allowed large-angle solution.  The theory errors are
significant but not dominant.  It should be emphasized that one must
include the correlation between the theoretical uncertainties in the
various experiments\footnote{A recent calculation \protect\cite{GKW}
which ignored the correlations obtained unrealistically large
theoretical uncertainties.}. For example, whatever value $T_c$ takes,
it is the same for all experiments.

Another technical comment concerns how one estimates confidence level
contours.  We have used the $\chi^2$ method.  However, the
$\chi^2$ method can only be rigorously justified
for error regions that are
Gaussian in the parameters, and it is difficult to interpret
when there is more than one allowed region.  For this
reason the interpretation in terms of confidence
levels should only be considered qualitative.  Another possibility is
to simply overlap the allowed regions (at a given confidence level) for
the different experiments~\cite{18a}.  This method has two drawbacks.
In some circumstances ({\it e.g.}, when two allowed bands cross) it
underestimates the true uncertainties.  On the other hand, it
sometimes admits solutions which are only marginally compatible with
each of several experiments and which are therefore really quite bad
fits.

For oscillations into active neutrinos $(\nu_\mu$ or $\nu_\tau)$ there
are two solutions allowed by all the data, the non-adiabatic (small
mixing angle) and the large-angle solutions,\footnote{The large-angle
solution would extend down to $\Delta m^2 \sim 10^{-7} \ev^2$ if one used
the overlap method \cite{18a}, but the $\chi^2$ in that region is very
large.} as can be seen in Figure~\ref{fig2}.
\begin{figure}
\vspace{5cm}
\caption{Allowed regions for MSW conversions of $\nu_e \ra \nu_\mu$ or
$\nu_\tau$, from \protect\cite{BHKL}.  The 90\% c.l. $(\Delta \chi^2 =
4.6)$ regions allowed by the Homestake, Kamiokande, and gallium
experiments and by the combined fit are shown.  The astrophysical and
nuclear uncertainties are included.  The corresponding allowed region
for oscillations into sterile neutrinos has a similar
small-angle non-adiabatic solution, but no large-angle solution.}
\label{fig2}
\end{figure}
The non-adiabatic solution gives a much better fit.  In this
region there is more suppression of the intermediate energy $\,^7Be$
neutrinos, accounting for the larger suppression seen by Homestake.
The large-angle fit is much poorer, corresponding to
$\chi^2= 3.8$ for $1 \;df$, because there the survival
probability varies slowly with neutrino energy.  One
can also consider the possibility that the $\nu_e$ is oscillating into
a sterile neutrino.  In this case the oscillation probabilities are
changed slightly due to the (small) neutron density in the sun.  A
much more significant effect is that the sterile neutrino cannot
undergo neutral current interactions in the Kamiokande experiment,
aggravating the difference between Homestake and Kamiokande.
The result is that there is no large-angle
solution\footnote{ The large angle solution for sterile neutrinos is
also most likely excluded by nucleosynthesis arguments \cite{sterile},
while  non-adiabatic parameters are allowed.} at 90\% C.L.  (it
only enters at 95\% confidence level).  There is a non-adiabatic
solution but even that yields a relatively poor fit $\chi^2 = 3.6$ for
1~$df$.

The conclusion is that MSW oscillations, combined with the SSM and
corresponding uncertainties, give an excellent description of the data,
especially the non-adiabatic solution for active neutrinos.  The
general range of $\Delta m^2$ is consistent with the expectations of
seesaw models motivated by grand unification.  However, the specific
regions of leptonic mixings are not in agreement with the naive
expectation $V_{\rm lepton} = V_{\rm CKM}$, which is predicted by some
of the simplest models.

\vglue 0.2cm
{\elevenit\noindent 4.2. MSW and Non-Standard Solar Model}
\vglue 0.1cm
It is also interesting to consider MSW
oscillations for an arbitrary core temperature $T_c$,
that is for NSSM.  (We will assume the same nuclear physics
uncertainties as in the SSM case.)  One now has three parameters, $T_c$,
$\sin^22\theta$, and $\Delta m^2$.  There are
sufficient constraints to determine all three \cite{BHKL}.  There is
an expanded non-adiabatic solution that corresponds to that in the
SSM.  In addition, the core temperature is determined; one finds $T_c
= 1.02^{+0.03}_{-0.05}$ at 90\%~C.L.  Similarly, there is a
large-angle solution with $T_c = 1.04^{+0.03}_{-0.04}$.  Thus the core
temperature is measured by the solar neutrino
experiments,\footnote{Of course, one of the original motivations for
the solar neutrino experiments was to probe the core of the sun.} even
allowing for the complication of MSW oscillations.  In particular, it
is consistent with the standard solar model prediction $T_c = 1 \pm
0.0057$, although at present the solar neutrino data only allows a
precision of $ \sim 5\%$.  Allowing for an arbitrary $T_c$ the MSW
parameter range broadens \cite{BHKL}.  For lower $T_c$ one obtains
smaller $\sin^22\theta$ for the non-adiabatic solution, while the
large angle solution disappears.  For larger temperatures the regions
move closer to each other, and the two regions are even connected by a
narrow neck at the 95\% C.L.

\vglue 0.2cm
{\elevenit\noindent 4.3. Future}
\vglue 0.1cm
In the future one will want to verify the MSW, uniquely determine
which solution is relevant, distinguish between conversions into
active or sterile neutrinos, {\it etc}.  To do this one needs better
statistics in the gallium experiments (as well as calibrations), and
new experiments with high event rates.  The SNO experiment
should observe spectral distortions if the non-adiabatic solution is
correct, measure the ratio of neutral current to charged current
($NC/CC$) events, and could even observe oscillations into
$\bar{\nu}_e$, which is not predicted by the SSM or MSW. The
SuperKamiokande experiment should also be able to observe
spectral distortions, while Borexino should be sensitive to the
$\,^7Be$ line and possible time dependence.  There may also be an
ICARUS experiment sensitive to the spectrum and some $NC/CC$.  The
implications of these are that a spectral distortion could essentially
unambiguously identify non-adiabatic MSW conversions~\cite{spectral}.
Observations of the $\,^7Be$ line are sensitive to nonstandard
neutrino properties.  The $NC/CC$ ratio distinguishes active neutrinos
from both sterile neutrinos and astrophysical solutions, and
for active neutrinos the $NC$ rate would give an independent measure of
$T_c$.
It is important to look for seasonal variations (expected from vacuum
oscillations), day/night effects (relevant to some parameter regions
of MSW), and correlations with the sunspot cycle (characteristic of
magnetic moments).  It is also useful to look for $\bar{\nu}_e$, which
would be indicative of neutrino decay or spin-flavor oscillations.

\vglue 0.6cm
{\elevenbf\noindent 5. Atmospheric Neutrinos}
\vglue 0.2cm
The predicted fluxes ${\nu_\mu}$ and ${\nu_e}$ produced by the
interactions of cosmic rays in the atmosphere are uncertain by around
20\%.  However, the ratio ${\nu_\mu}/{\nu_e}$ is believed to be
accurate to $\sim$~5\% \cite{penn}.  There are additional
uncertainties associated with interaction cross-sections, particle
identification, {\it etc}. The Kamiokande and IMB groups have observed
a deficit in the ratio of contained muon and electron events
\beq \frac{(\mu/e) |_{\rm data} }{ (\mu/e) |_{\rm
theory} } = \left\{ \begin{array}{ccc} 0.65 \pm 0.08 \pm 0.06 & ,
& {\rm Kamiokande\ \protect\cite{8a}} \\ 0.54 \pm 0.05 \pm 0.12 & . &
{\rm IMB\ \protect\cite{imb}} \end{array} \right. \eeq
This effect, if real, suggests the possibility of $\nu_\mu \ra
\nu_\tau$ or possibly $\nu_\mu \ra \nu_e$.  It probably is not
compatible with sterile neutrino oscillations $\nu_\mu \ra \nu_s$,
because for the relevant parameter range one would have a clear
violation of the nucleosynthesis bound \cite{sterile}.  From
Figure~\ref{fig1} one sees that the oscillation hypothesis\footnote{ An
interesting alternative involves an
enhancement in the number of positrons due to proton decay $p \ra e^+
\nu \bar{\nu}$ in the detector \protect\cite{WKL}.  However, this is
disfavored by the observed spectrum.} requires a mass range $\Delta
m^2 \sim (10^{-3} {-1}) \ev^2$, larger than that relevant to the solar
neutrinos, and large mixing angles such as $\sin^22 \theta \sim 0.5$.

On the other hand, the IMB collaboration \cite{imb} has measured the
flux of upward-going muons, and has concluded from the nonobservation
of any deficit that most of the range suggested by the contained
events is excluded.  However, the upward-going rate is sensitive to
the absolute calculation of the flux.  Moreover, it is sensitive to
the deep inelastic cross-section needed to produce the muon, and
recently it has been argued \cite{penn} that the Owens cross-sections
\cite{penna}, which give a better fit to the accelerator data than the
EHLQ \cite{penna} (used by IMB), lead to a weaker constraint
compatible with the contained events.  IMB also studied the ratio of
upward through-going to stopping muons, which is more reliable than
the absolute calculation; this excludes the low $\Delta m^2$ part of
the region suggested by the contained events, but allows most of the
range above $\Delta m^2 > 10^{-2} eV^2$.  The situation is clearly
confused.  Oscillations are suggested but certainly not proved, and
there are still many uncertainties associated with the fluxes,
cross-sections, and particle identification.

\vglue 0.6cm
{\elevenbf\noindent 6. Implications}
\vglue 0.2cm
It is hard to make concrete predictions for neutrino masses in most
models.  However, the $\Delta m^2$ range suggested by the solar
neutrinos is compatible with the general range expected in
quadratic up-type seesaw models, such as in grand unified theories
\cite{BHKL}, for which $m_{\nu_i} \sim m_{u_i}^2/M_N$,
where $u_i = u,c,t$ and $M_N$ is the heavy neutrino mass.  For $M_N
\sim 10^{11} - 10^{16}$~GeV one obtains the appropriate mass range,
for oscillations into $\nu_\tau$ for $M_N\sim 10^{16}$ GeV, and into
$\nu_\mu$ for $M_N \sim 10^{11}$~GeV.  However, the simplest models
predict equal lepton and quark mixing angles, $V_{\rm lepton} = V_{\rm
CKM}$, which is not satisfied by the data unless $T_c$ is far from the
SSM \cite{BHKL}.\footnote{There may a broader class of solutions if
one allows for three-neutrino oscillations \protect\cite{HKP}.}

The various hints suggest two general scenarios.  One could
have $\nu_e \ra \nu_\mu$ in the sun for $m_{\nu_e} \ll m_{\nu_\mu}
\sim 3 \x 10^{-3}$~eV, with $\nu_\tau$ a component of the dark matter
($m_{\nu_\tau} \sim few $~eV).  This pattern is compatible with the
mass predictions of GUT-type seesaws.  However, in this scenario there
is no room for oscillations to account for the atmospheric neutrinos.
A separate possibility is that again $\nu_e \ra \nu_\mu$ in the sun,
with $\nu_\mu \ra \nu_\tau$ oscillations with $m_{\nu_\tau} \sim (0.1 -
0.6)$~eV for the atmospheric neutrinos.  In this case there
would be no room for hot dark matter.  This second solution requires
large leptonic mixings, $\sin^2 2 \theta_{\nu_\mu \nu_\tau} \sim 0.5$.
This is certainly allowed, but is not what one would expect in most
models: if there is a large hierarchy of masses one typically expects
$\sin^22 \theta \ll 1$, while $\sin^2 2 \theta \simeq 1$ if the
neutrinos are nearly degenerate.

Given the constraints from LEP, nucleosynthesis, {\it etc}., it is
hard to find a reasonable scenario which would combine both solar and
atmospheric neutrino oscillations with HDM.  This is mainly because
there are just not enough neutrinos.  About the only viable scenario
is to invoke non-adiabatic oscillations into a sterile neutrino in the
sun, and to assume $m_{\nu_\mu} \sim m_{\nu_\tau} \sim {\rm few} $~eV
to provide hot dark matter.  One would need nearly
degenerate neutrinos so that $\Delta m^2 = m_{\nu_\tau}^2 - m_{\nu_\mu}^2
\sim 10^{-1} {\rm eV}^2$ for the atmospheric neutrino oscillations.
This pattern is possible but seems very contrived.  Of course, if one
relaxes some of the experimental constraints the range of
possibilities increases.  For example, if one ignored the Homestake
experiment one could have $\nu_e \leftrightarrow \nu_\mu$ for both
solar and atmospheric neutrinos, with $\Delta m^2 \sim 10^{-2} {\rm
eV}^2$.  However, in this case much of the rationale for a
nonastrophysical solar solution would also disappear.

\vglue 0.6cm
{\elevenbf\noindent 7. Viable Models
Inspired by Coupling Constant Unification}
\vglue 0.2cm
There has been a recent revival of interest in grand unification,
inspired by the success of the unification of the coupling constants
\cite{33a} in the supersymmetric extension of the standard model.  The
various models have implications for neutrino mass, and predict the
right general mass range for the MSW solution.  However, the simplest
models do not agree with the data in detail, at least not assuming the
SSM.

Let us first consider the {\it old} supersymmetric grand unified
theories \cite{BHKL}.  These are the type of models that were popular
ten years ago, which allowed large and complicated Higgs
representations, such as $SO_{10}$ models with both 10 and
126-dimensional Higgs multiplets.  If only the $\varphi_{10}$
generates the masses for the quarks and charged leptons, then one has
the simple relation $m_D = m_u$, $m_e = m_d$ for the mass matrices at
the large scale, where $m_D$ is the matrix of Dirac neutrino masses.
If one also introduces a $\varphi_{126}$ whose only role is to
generate a large majorana mass for the new right-handed neutrinos,
then one obtains typical see-saw predictions
\beqa m_{\nu_i} & \sim & c_i \frac{m_{u_i}^2}{M_{N_i}} \nonumber \\
m_{N_i} &\sim & (10^{-2} - 1) M_X, \eeqa
where $c_i \sim 0.05 - 0.1$ is a radiative correction and $M_X \sim
10^{16}$ GeV is the unification scale \cite{BHKL,33a}.  For these
assumptions
\beqa m_{\nu_e} & \leq& 10^{-11}\; \eV \nonumber \\
m_{\nu_\mu} &\sim& (10^{-8} - 10^{-6})\; \ev \nonumber \\
m_{\nu_\tau} &\sim& (10^{-4} - 1) \;\ev \eeqa
This is the correct mass range for $\nu_e \ra \nu_\tau$ in the sun.
However, for a wide range of parameters one also expects
\beq V_{\rm lepton} \sim V_{\rm CKM}, \label{eq5} \eeq
implying a smaller mixing angle than is favored by the data.  A more
serious problem is that the same models predict
$m_d/m_s = m_e/m_\mu$, which fails by an order of magnitude
\cite{33b}.

Within the same class of theories there have been more sophisticated
recent models in which the $\varphi_{126}$ not only generates
$M_{N_i}$ but also contributes to the quark, charged lepton, and
(possibly) Dirac neutrino masses.  For example, Dimopoulos, Hall, and
Raby \cite{DHL} have revived an interesting model of Georgi and
Jarlskog \cite{GJ}, in which both $\varphi_{10}$ and $\varphi_{126}$
contribute and for which, in addition, there are many discrete
symmetries.  These are arranged so that there are a number of zeros in
the mass matrices, and so that each non-zero element is generated by a
unique representation.  One can then fix up the problematic mass
relation, replacing it by the acceptable $9 m_e/ m_\mu = m_d/m_s$.
The model leads to interesting predictions for the CKM matrix and
$m_t$.  The neutrino mixing angle predictions are somewhat modified
with respect to (\ref{eq5}), but are still not in perfect agreement
with the data unless $T_c > 1$.  Other interesting models and
calculations are in \cite{BM,BS,FTY,ELN}.

One can also consider the {\it new} supersymmetric grand unified
theories inspired by superstrings.  These differ in that most
superstring compactifications do not lead to large Higgs
representations.  In particular, in $SO_{10}$ there is no
$\varphi_{126}$.  In such models there is no way to generate
$m_{N_i}$ at tree-level and it is not obvious how to even have a
seesaw.  However, the relevant right-handed neutrino masses could be
generated by effective non-renormalizable operators\footnote{For
another use of effective operators, see \protect\cite{42a}.} left over
from the underlying theory \cite{NS}.  For example, one may
have an effective operator \cite{CL}
\beq L_{\rm eff} = \frac{C}{m_C} \bar{N}^c_L N_R \Phi \Phi, \eeq
where $m_C \sim 10^{18} \gev$ is the compactification scale and $\Phi$
is a 10-plet of $SO_{10}$.  For $\langle \Phi \rangle \sim M_X \sim
10^{16} \gev$ (the unification scale),
\beq M_N \sim C \frac{ | \langle \phi \rangle |^2}{m_C} \sim
10^{-2} C |\langle \Phi \rangle |.\eeq
An interesting example are large-radius Calabi-Yau spaces, for
which
\beq C \sim e^{-R^2/\alpha'} \;\; , \;\; \frac{M_X}{m_C} \sim
\frac{1}{\sqrt{2R^2/\alpha'}}, \eeq
where $R'$ is the radius and $\alpha'$ is the string tension, yielding
an intermediate scale $M_N \sim 10^{11} \gev$, and
\beq m_{\nu_e} \sim 10^{-7}\; \ev\;\;,\;\; m_{\nu_\mu} \sim 10^{-3}
\;\ev \;\;,\;\; m_{\nu_\tau} \sim 10 \;\ev. \eeq
One could then have $\nu_e \ra \nu_\mu$ in the sun and the $\nu_\tau$
would be a candidate for the HDM.  There is no quantitative prediction
for the leptonic mixing angles.

The crucial feature of this type of seesaw is the intermediate scale
$m_N \sim 10^{11} \gev$, considerably lower than a GUT scale.  This
may also come about in other ways.  For example, in ordinary
(non-supersymmetric) grand unified theories it is possible that there
are intermediate scales associated with stages of the symmetry
breaking.  For an intermediate scale in this range one would typically
reproduce the above predictions for the neutrino mass \cite{BHKL}.
However, in the simplest versions one might again have trouble with
the prediction $V_{\rm lepton } \sim V_{\rm CKM}$.

\vglue 0.6cm
{\elevenbf\noindent 8. Summary/Conclusions}
\vglue 0.2cm
The relative rates for the various solar neutrino experiments are more
significant than the fact that they are all smaller than the standard
solar model prediction, and strongly suggest that astrophysics is not
the solution to the solar neutrino problem (unless, of course, one of
the experiments is wrong).  There are various MSW solutions that agree
with the data.  Assuming the standard solar model for the initial
production, there are two MSW solutions, non-adiabatic and large-angle
for oscillations into active neutrinos, with the non-adiabatic
favored.  For oscillations into sterile neutrinos there is only a
non-adiabatic solution.  One can also allow the possibility of
nonstandard solar models.  In a three-parameter fit one obtains $T_C
\sim 1.02^{+0.03}_{-0.05}$, {\it i.e.}, the core temperature is
determined to within 5\% even allowing for MSW oscillations, and
agrees with the predictions of the standard model.

The deficit observed in the ratio $\nu_\mu/\nu_e$ as measured in
contained atmospheric neutrino events suggests $\nu_\mu \ra \nu_\tau$
oscillations, or possibly $\nu_\mu \ra \nu_e$, but not oscillations
into sterile neutrinos.  However, the interpretation is still clouded
by uncertainties in particle ID, flux, and cross-sections.

The $\Delta m^2$ range suggested by the solar neutrinos is compatible
with the GUT seesaw.  However, $V_{\rm lepton} \neq V_{\rm CKM}$
unless the temperature $T_c$ is far from the standard solar model.
There is still room for the $\nu_\tau$ to play the role of hot dark
matter.  One can have simple solutions involving $\nu_e \ra \nu_\mu$
in the sun and $\nu_\tau$ as the HDM, or, alternately, $\nu_e \ra \nu_\mu$ in
the sun and $\nu_\mu \ra \nu_\tau$ to account for atmospheric
neutrinos.  However, it is hard to combine all of these given other
constraints, except in a complicated scenario which seems rather
contrived.

Clearly, we need more and better solar neutrino experiments, with
better statistics, measurements of the spectrum, $NC/CC$ ratio, and
the $\,^7Be$ line, searches for $\bar{\nu}_e$, and a resolution of
whether there is any time dependence.  Finally, it is important to
continue the program of laboratory experiments.  In particular,
experiments searching for $\nu_\mu \ra \nu_\tau$ for small $\sin^2 2
\theta$ are strongly indicated by the hot dark matter scenario, while
$\nu_\mu \ra \nu_\tau$ and $\nu_e \ra \nu_\tau$ searches for small
$\Delta m^2$ and large angles are motivated by the atmospheric
neutrinos.


\begin{thebibliography}{000}
\bibitem{penna} References may be found in \cite{penn}.
\end{thebibliography}
\end{document}